\documentclass[journal]{IEEEtran}
\usepackage{}

\ifCLASSINFOpdf
\else
\fi
\usepackage{array}

\usepackage[cmex10]{amsmath}
\usepackage{bm}
\usepackage{enumerate}
\usepackage{amsfonts}
\usepackage{graphicx}
\usepackage{cite}
\usepackage{booktabs}
\usepackage{multirow}
\usepackage[linewidth=0.5pt]{mdframed}
\usepackage{threeparttable}
\usepackage{fixltx2e}
\usepackage{color}
\usepackage{xcolor}
\usepackage{subeqnarray}
\usepackage{cases}
\usepackage{makecell}
\usepackage{algorithmic}
\usepackage{algorithm}
\usepackage{bbold}
\usepackage{amssymb}
\usepackage{epstopdf}
\usepackage{amsthm}
\usepackage{algorithmic}
\usepackage{algorithm}
\theoremstyle{definition}

{ \bgroup
  \addtolength\abovedisplayshortskip{#1}
  \addtolength\abovedisplayskip{#1}
  \addtolength\belowdisplayshortskip{#1}
  \addtolength\belowdisplayskip{#1}}
{\egroup\ignorespacesafterend}
\ifCLASSOPTIONcompsoc
\usepackage[caption=false,font=normalsize,labelfon
t=sf,textfont=sf]{subfig}
\else
\usepackage[caption=false,font=footnotesize]{subfig}

\fi

\begin{document}
\bibliographystyle{IEEEtran}

%-----------------------------------------
%-----------------------------------------
%-----------------------------------------

% \title{Guiding Network Cascading Failure Search with an Interpretable Graph Learning Approach}
\title{Guiding Cascading Failure Search with Interpretable Graph Convolutional Network}
% \title{Graph Convolutional Network Guided\\Interpretable Cascading Failure Search}
% 能否修改题目，加入machine learning关键词，相比较之下，data-driven用烂了
\author{
Yuxiao~Liu,~\IEEEmembership{Student Member,~IEEE,}
Ning~Zhang,~\IEEEmembership{Senior Member,~IEEE,}
Dan~Wu,~\IEEEmembership{Member,~IEEE,}
Audun~Botterud,~\IEEEmembership{Member,~IEEE,}
Rui~Yao,~\IEEEmembership{Member,~IEEE,}
and~Chongqing~Kang,~\IEEEmembership{Fellow,~IEEE}

\thanks{ This work was supported in part by the National Science Foundation of China under Grant U1766212, and in part by the Tsinghua University Initiative Scientific Research Program 20193080026. (\emph{Corresponding authors: Ning Zhang})
  
Yuxiao Liu, Ning Zhang and Chongqing Kang are with the Department of Electrical Engineering, Tsinghua University, Beijing 100084, China. (liuyuxiao16@mails.tsinghua.edu.cn; ningzhang@tsinghua.edu.cn; cqkang@tsinghua.edu.cn)

Dan Wu, Audun Botterud are with the Laboratory of Information and Decision Systems, Massachusetts Institute of Technology, Cambridge, MA 02139 USA. (danwumit@mit.edu; audunb@mit.edu)

Rui Yao is with the Argonne National Laboratory, Lemont, IL 60439, USA (yaorui.thu@gmail.com). Rui Yao is supported by DOE office of science under contract DE-AC02-06CH11357. 
}% <-this % stops a space
% % %\thanks{This work was supported by the National Natural Science Foundation of China (No. 51325702,51307092)}% <-this % stops a space

}

\maketitle

%-----------------------------------------
%-----------------------------------------
%-----------------------------------------

\begin{abstract}
% 一个非常重要的问题就是你都不是在突出解决了cascading什么问题，都在说你的GCN多好，比如没有数据驱动模型就不行！！
Power system cascading failures become more time variant and complex because of the increasing network interconnection and higher renewable energy penetration.
High computational cost is the main obstacle for a more frequent online cascading failure search, which is essential to improve system security.
In this work, we show that the complex mechanism of cascading failures can be well captured by training a graph convolutional network (GCN) offline. 
Subsequently, the search of cascading failures can be significantly accelerated with the aid of the trained GCN model.
We link the power network topology with the structure of the GCN, yielding a smaller parameter space to learn the complex mechanism.
We further enable the interpretability of the GCN model by a layer-wise relevance propagation (LRP) algorithm.
The proposed method is tested on both the IEEE RTS-79 test system and China's Henan Province power system. 
The results show that the GCN guided method can not only accelerate the search of cascading failures, but also reveal the reasons for predicting the potential cascading failures.
% 其实最后一点真的非常重要，因为你通过模型的方法也很难搞定
\end{abstract}
% Note that keywords are not normally used for peer review papers.

\IEEEpeerreviewmaketitle

%-----------------------------------------
%-----------------------------------------
%-----------------------------------------

\section{Introduction}
In interconnected power systems, some local failures can yield subsequent failures which may eventually lead to a severe damage to the system. 
This sequence of failures is referred to as the ``cascading failure''.
Searching the critical cascading failure paths is of great importance for securing the operation conditions.
Once some critical cascading failure paths are detected, various preventive strategies can be applied, such as re-dispatching the generators to decrease the power flow of high risk branches, strengthening the transmission lines inspection, increasing the generation reserves, etc~\cite{vaiman2012risk,bienstock2015electrical}.
With the increasing interconnection of power networks and the penetration of stochastic renewable generation, cascading failure paths become more time variant and complex~\cite{haes2019survey,athari2017impacts}.
Therefore, the cascading failure search calls for a more frequent online calculation to safeguard a reliable operation of the power system.
However, the high computational cost, from the ``curse of combinatorial dimensionality'' of the contingencies, is the major obstacle towards online cascading failure search.
For a power system with $N$ components, the number of $N-k$ contingencies is $N!/(N-k)!$, if the sequence matters~\cite{vaiman2012risk}.

Efforts have been made to the efficient analysis of cascading failures.
One approach is to modify the random sampling strategies of components' failures, so that more severe paths could be detected within less samples.
Such techniques include the importance sampling method~\cite{chen2013composite,henneaux2014improving}, the splitting method~\cite{kim2012splitting,wang2014efficient}, the random chemistry method~\cite{eppstein2012random}, etc.
Although these methods can significantly improve the sampling efficiency compared to random sampling, they still suffer from duplicated simulations of the same cascading paths.
To avoid duplicated simulations, some methods model the cascading failures simulation as a Markov chain search problem.
Yao \textsl{et al.} proposed the risk estimation index to quantify the priority of each search attempt~\cite{yao2016risk}.
Liu \textsl{et al.} accelerated the cascading failure simulation by simplifying multiple related DC optimal power flow problems into affine calculation problems~\cite{liu2019fast}.
Nonetheless, the Markov chain search still suffers from high computational complexity from the large search space of network cascading failures.
Soltan \textsl{et al.} analytically computed the disturbance value (or the redistribution of power flow) caused by a failure to replace the power flow calculation~\cite{soltan2015analysis,soltan2017analyzing}.
Such methods provide an efficient way for contingency analysis, but they cannot well capture the islanding or re-dispatch events in the whole cascading failure procedure.
Because of the complex mechanism of cascading failures, it still remains a challenging problem to effectively search for the cascading failure paths that result in load shedding.

Recently, some machine learning models are proposed to capture the complex mechanism of cascading failures.
% Recently, some work capture the complex mechanism of cascading failures by training machine learning models with a large number of simulation data.
The models learn the mapping from simulated power system operational data to the vulnerability of the components, thereby accelerating the online cascading failure search.
Several different machine learning models are investigated recently, such as the deep neural network (DNN)~\cite{li2018alphago}, the convolutional neural network (CNN)~\cite{du2019achieving,arteaga2019deep}, the auto encoder~\cite{sun2018deep}, the Q-learning model~\cite{yan2016q,zhang2019online}, and the influence graph model~\cite{hines2016cascading}.
Some of the aforementioned methods train the machine learning model to directly predict whether the systems are stable or not under certain contingencies~\cite{li2018alphago,du2019achieving,sun2018deep,hines2016cascading}; while others only use the machine learning model to guide the search with the physical models (e.g. the power flow based model)~\cite{yan2016q,zhang2019online}.
% Because machine learning model can only provide an approximation of the real world, we choose the latter approach for more reliable results.
Two vital problems remain to be addressed to build better machine learning models.
One challenge is that the number of model parameters increases significantly with the size of the power system, making the training process time-consuming and prone to over fitting.
Another challenge is the lack of interpretability of machine learning models that hinders the practical applications in power systems~\cite{cremer2019optimization}. 
% Secondly, the lack of interpretability is the main challenge for the wide adoption of machine learning models in power system applications~\cite{cremer2019optimization}.
It is necessary for the power system operators to check the logic of the models and to understand the underlying factors that cause cascading failures.  % not only efficiently identify the risk levels but also

To address the aforementioned problems, we propose a graph convolutional network (GCN)~\cite{wu2019comprehensive,du2017topology} based approach for an efficient online search of cascading failures.
The GCN model takes advantage of the fact that the mechanism of cascading failures is highly related to the topology of the power system, e.g., the failure of a branch will more likely result in the load shedding of the nearby buses, or the increase of a bus load will more likely trigger the protection relays of the nearby branches, etc.
The proposed GCN approach uses graphical model to learn this mechanism and theoretically requires far less parameters than other machine learning models (DNN or CNN)~\cite{li2018alphago,arteaga2019deep}. % less parameters有点问题，并不是“需要”更少的参数，而是能够用更少的参数表征复杂mechanism，是否还要说出efficiency
To increase the interpretability of the model, we tailor the model inputs into four types of factors: the topology of the network, the impact of the protection relays, the branch flow, and the amount of load consumption.
Furthermore, we explain the GCN model by a layer-wise relevance propagation (LRP) algorithm~\cite{baldassarre2019explainability} and quantify the contribution factors of model predictions.
To the best of the authors' knowledge, the paper is the first to implement the interpretability of GCN models in power grid analysis.
The proposed algorithms can identify the contributing factors that cause the cascading failures, such results could help manage the power system to mitigate the damage.
The contributions of this work are as follows:
\begin{enumerate}[1)]
  \item We propose a GCN based approach to guide the search of cascading failure for a more efficient detection of load shedding events, which enables an online deployment of cascading failure search to better protect the power system.
  \item The proposed GCN model links the power network topology with the structure of neural networks and can learn the complex power system cascading failure mechanism within a smaller model parameter space.
  \item We implement the LRP algorithm to enable the interpretability of the GCN model so as to narrow down the contributing factors to the cascading failures. % We design the model inputs as four types of factors with physical interpretations (i.e. topology, protection relay, branch flow, and bus load) and narrow down the causes of the potential cascading failures.
\end{enumerate}

The remainder of this paper is organized as follows.
Section~\ref{section_cascading_failures} introduces the power system cascading failure simulation and search strategy used in this work.
In Section~\ref{section_GCN}, we propose the GCN model and its interpretation method for cascading failure search.
Thereafter, the case study is presented in Section~\ref{section_case_studies}.
Finally, Section~\ref{section_conclusions} draws the conclusions.

% 先把整体框架说一下
%look ahead search??
% 是否需要说GCN的可视化是一个非常前沿的方向？？
% this work focus on branch failures, as in ...  do
%  realtime在后面说，this paper aims at finding more in less attempts of contingencies... so that 我们可以尽快行动？

%-----------------------------------------
%-----------------------------------------
%-----------------------------------------

\section{Power System Cascading Failure Search}
\label{section_cascading_failures}
\subsection{The Framework of Cascading Failure Search}

\begin{figure}[htb!]
	\centering
		\includegraphics[width=1.0\linewidth]{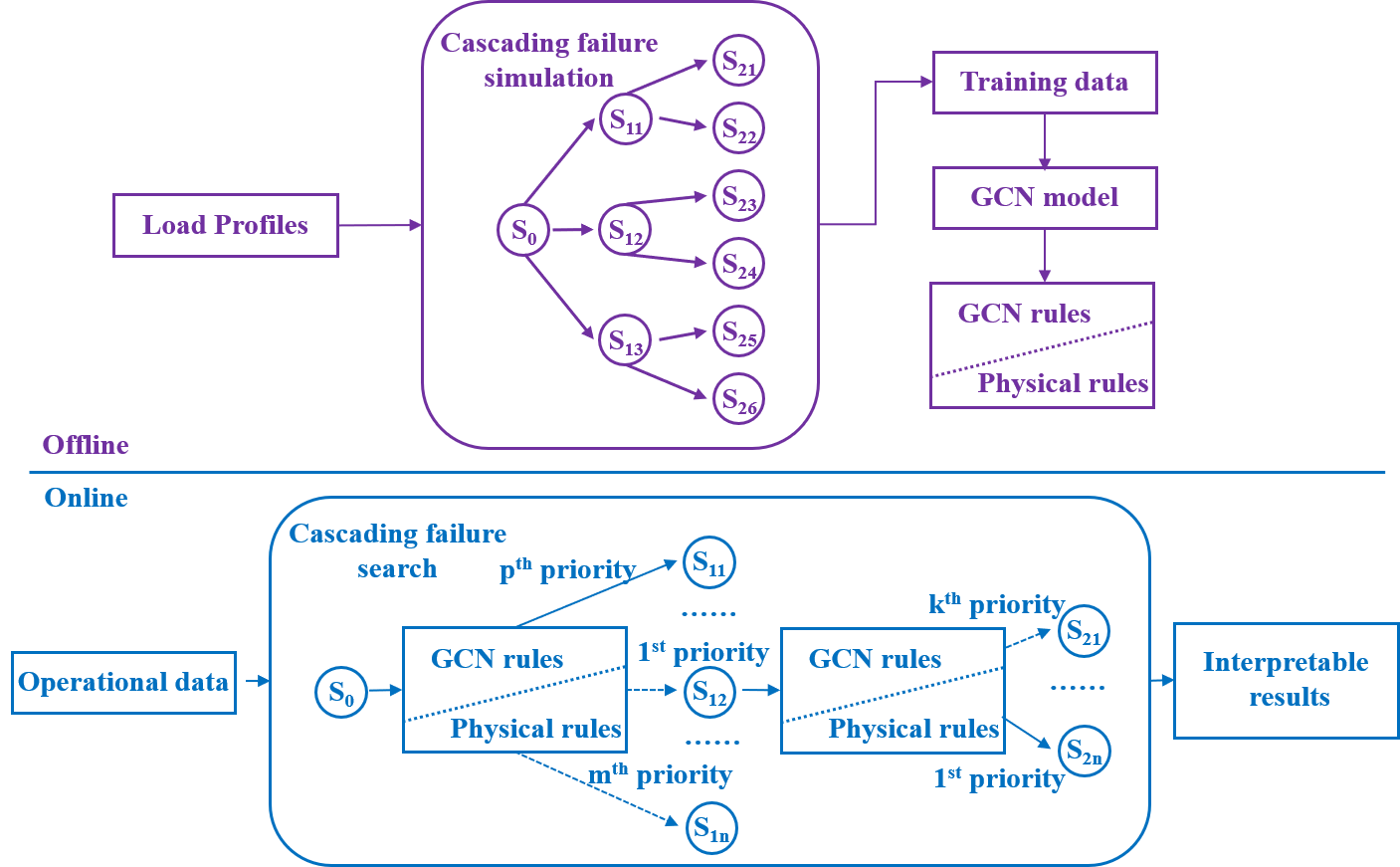}
	\caption{The framework of power system cascading failure search over different states, $\mathbf{S}$.}
	\label{fig_framework}
\end{figure}

The framework of power system cascading failure search is shown in \figurename~\ref{fig_framework}.
The purpose of this work is to figure out more cascading paths that result in load shedding in a certain number of online search attempts.
Firstly, we generate the power system cascading failure path offline as training samples.
We then train a GCN model that maps from the current operational state to the vulnerabilities of each component.
We combine the trained GCN model and selected physical rules as the guidance in the online cascading failure search, where the components with high vulnerabilities are searched with priorities.
The interpretable results of why certain components are predicted with high vulnerabilities are also provided by the method.
\subsection{Power System Cascading Failures Simulation}
\label{section_OPA}
In this work, we simulate the power system cascading failures based on the ORNL-PSerc-Alaska (OPA)~\cite{carreras2004complex,ren2008long} model, which is widely adopted and developed in~\cite{mei2009improved,kim2012splitting,yao2016risk,zhang2019online,liu2019fast}.
The OPA model simulates the cascading failures triggered by branch failures in the power system.
Two types of failures are considered.
One type is caused by random events such as sagging and tree contact.
The other type is triggered by the action of protection relays once the branch flow exceeds the preset threshold.
In this work, we simulate a limited number of random failures (the first type), while we simulate all the protection relay actions once they happen.
The OPA model applied in this work can be summarized as follows.
\begin{enumerate}[1)]
  \item \emph{Step 1:} Set a positive integer value $R$, so that we only consider the cases of no more than $R$ random failures. Run the DC optimal power flow (DCOPF) given the initial load profile.
  \item \emph{Step 2:} If the number of random branch failures is less than the preset value $R$, disconnect a branch. Else, end the simulation.
  \item \emph{Step 3:} Run the DC power flow (DCPF). Disconnect the branch that triggers the protection relays and run \emph{Step~3} again. If the network separates into islands, conduct load shedding for the islands which have a lack of generation.
  \item \emph{Step 4:} Re-dispatch the system with the objective to minimize the total load shedding and with the constraints of the system security rules (e.g. the branch flow constraints). Run \emph{Step~2}.
\end{enumerate}
Note that the OPA model is implemented as an indicative test bed, which is not the main focus of this work.
We refer to~\cite{carreras2004complex,ren2008long} for more details of the OPA model.
% The proposed cascading failure search method is general to the simulation models.

\subsection{Branch Vulnerabilities to Guide the Search}
When simulating the cascading failures, the major computational burden comes from the combinatorial explosion in \emph{Step 2}, i.e., we have to disconnect every branch that may fail and simulate every possible case.
To ease the computational burden, we train a model that guide the search in \emph{Step 2}.
That is, we provide a priority on the search order at \emph{Step 2}, given the current operational state.
To this end, we organize the data into input $\bm{X}^{GCN}$ and output $\bm{y}^{GCN}$ to train the GCN model.
\begin{equation}\label{eq_GCN}
  \bm{y}^{GCN}=f(\bm{X}^{GCN})
\end{equation}
The input is the current operational state and the output is a boolean vector that quantifies the vulnerability of the branches.
The $k$th variable of the vector $y^{GCN}_k$ is a binary value representing whether the load shedding occurs after disconnecting branch $k$ (0-without load shedding, 1-load shedding).

Afterwards, we adopt some physical rules to assist the GCN model.
We calculate the value of line outage distribution factors (LODF)~\cite{guler2007generalized,tejada2017security}. 
For a system with $N$ buses, we have:
\begin{equation}\label{eq_LODF}
  D_{m}^{k}=
  \frac{X_{m}^{k}/x_{m}}
  {1-X_{k}^{k}/x_{k}},
  \text{ }
  X_{m}^{k}=\mathbf{M_m^T}\mathbf{X}\mathbf{M_k},
\end{equation}
where $D_{m}^{k}$ denotes the additional branch flow in branch $m$ (as a fraction of the initial branch flow in branch $k$) after the outage of branch $k$, $x_{k}$ is the branch reactance of branch $k$, $\mathbf{X}$ is the $N\times N$ nodal reactance matrix, and $\mathbf{M_k}$ is a $N\times 1$ vector for branch $k$ from bus $i$ to bus $j$:
\begin{equation}\label{eq_M}
  \mathbf{M_k} = {{\left[ \underset{1}{\mathop{0}}\,\text{ }...\text{ }\underset{i}{\mathop{\text{1}}}\,\text{ }...\text{ }\underset{j}{\mathop{\text{-1}}}\,\text{ }...\text{ }\underset{N}{\mathop{\text{0}}}\, \right]}^{T}}.
\end{equation}
After the outage of branch $k$, the vulnerability of branch $k$ can be quantified by the total branch flow in branch $m$ as a fraction of its long term maximum branch flow $L_m^{max}$:
\begin{equation}\label{eq_Vulnerability}
  {\alpha}_m^k = |L_m+D_m^kL_k|/L_m^{max}
\end{equation}
When ${\alpha}_m^k$ exceeds the protection relay threshold $\beta$, the protection relay will disconnect the branch $m$.
When the system separates into islands, the denominator of~\eqref{eq_LODF} ($1-X_{k}^{k}/x_{k}$) is~0.
In this case, we set $\alpha_m^k$ to another value to avoid numerical problem.
We set $\alpha_m^k=\beta$ because $\beta$ already reaches the threshold to trigger the protection relays after the power flow redistribution, which is large enough to represent the vulnerability of branch $k$.
\begin{equation}
  \begin{aligned}\label{eq_LODF'}
    \hat{\alpha}_{m}^{k}=\left\{ \begin{matrix}
      {\alpha}_{m}^{k} & 1-X_{k}^{k}/x_{k} \neq 0  \\
      \beta & 1-X_{k}^{k}/x_{k}=0  \\
   \end{matrix} \right.     
  \end{aligned}.
\end{equation}
% For system islanding, one can set $\hat{\alpha}_{m}^{k}$ as any value larger than $\beta$ to represent the vulnerability of this branch.
% But $\beta$ already reach the threshold to trigger the protection relays and the branch disconnections, which is large enough to represent the priority to disconnect the corresponding branch.
Then, we define the vulnerability vector based on physical rules as $\bm{y}^P$, with the $k$th variable as:
\begin{equation}\label{eq_vul_physical}
  y_k^P=\underset{m}{\mathop{\max}}\hat{\alpha}_{m}^{k}.
\end{equation}
\eqref{eq_LODF'}-\eqref{eq_vul_physical} denotes that the protection relay (of branch $m$ with the largest $\hat{\alpha}_{m}^{k}$) will take action or the system will separate once $y_k^P$ exceeds $\beta$.
Larger $y_k^P$ intuitively implies the failure of branch $k$ maybe more likely to incur load shedding.
Hence, $y_k^P$ serves as a fast yet not accurate factor to quantify the branch vulnerabilities.
Finally, we guide the online search of cascading failures by both the vulnerabilities learned by the GCN $\bm{y}^{GCN}$ and the vulnerabilities from physical rules $\bm{y}^{P}$.

\subsection{The Online Search Algorithm}
\begin{algorithm}[ht]
  \caption{The Online Search Algorithm}
  \label{algorithm_online_search}
\begin{algorithmic}[1]
  \STATE {\textbf{Step 1*:} Run the DCOPF given the initial load profile;}
  \STATE {\textbf{Step 2*:} Use the GCN and physical rules to guide the search;
    \WHILE{the number of random branch failures is less than a preset value $R$,} 
      \STATE {calculate $\bm{y}^{GCN}$ with~\eqref{eq_GCN}}; 
      \FORALL{$y_k^{GCN}=1$,} 
        \STATE{disconnect the branch $k$ and run the \emph{Step~3} and \emph{Step~4} of the OPA model in Section~\ref{section_OPA}; run \textbf{Step~2*}} to simulate the next failure; 
      \ENDFOR 
      \STATE {calculate $\bm{y}^{P}$ with~\eqref{eq_LODF}-\eqref{eq_vul_physical}}; 
      \FORALL{$y_k^{P}$ from the largest to the smallest,} 
        \STATE{disconnect the branch $k$ and run the \emph{Step~3} and \emph{Step~4} of the OPA model in Section~\ref{section_OPA}; run \textbf{Step~2*}} to simulate the next failure; 
      \ENDFOR 
    \ENDWHILE}
  \STATE {\textbf{Step 3*:} Output all the cascading paths that cause load shedding.}
\end{algorithmic}
\end{algorithm}

The online search model only differs from the OPA model in the term of the order to disconnect the branches in \emph{Step~2}, so that we can detect the cascading failures that result in load shedding more efficiently.
That is, we calculate the $\bm{y}^{GCN}$ and $\bm{y}^P$ before running \emph{Step~2}.
Every time we run \emph{Step~2}, we first disconnect the branch that $y_k^{GCN}=1$, then we disconnect the branch from the largest $y_k^P$ to the smallest $y_k^P$.
The aforementioned process is formulated in \textbf{Algorithm~\ref{algorithm_online_search}}.

% 这里也可以提一下我们的特点是单步，因此可以具有更好的interpretability
% 最后还要说明我们是否要穷尽所有的search，说明我们是要在更少的attempt中找到更多的fail
\section{Graph Convolutional Network}
\label{section_GCN}
\subsection{Learning on Graphs}

\begin{figure}[ht]
	\centering
	\begin{minipage}{2.7cm}
	\centering
	\includegraphics[height=0.9in]{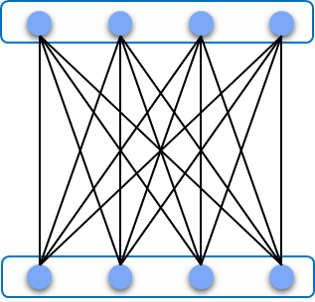}
	\centerline{\footnotesize{(a)}}
	\end{minipage}
	\begin{minipage}{2.7cm}
	\centering		
	\includegraphics[height=0.9in]{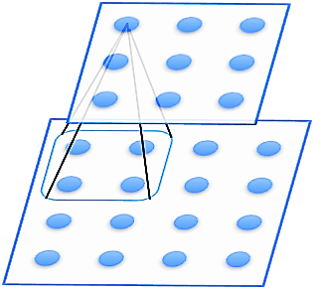}
	\centerline{\footnotesize{(b)}}
  \end{minipage}
  \begin{minipage}{2.8cm}
    \centering		
    \includegraphics[height=0.9in]{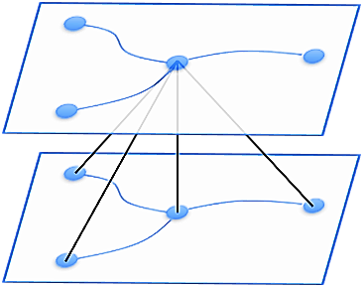}
    \centerline{\footnotesize{(c)}}
    \end{minipage}
  \caption{A demonstration of three types of layers. (a) The fully connected layer in DNN model. (b) The convolutional layer in CNN model. (c) The graph convolutional layer in GCN model.
  \label{fig_layers}
	}
\end{figure}

We focus on how to build the model that maps from the current operational state to the vulnerability of the branches, as formulated in~\eqref{eq_GCN}.
The operational data of the power system is physically represented and correlated in the form of graphs.
The mechanism of cascading failures is also closely related to the graphical structure of the system.
Therefore we choose the GCN model over the DNN model~\cite{li2018alphago} and the CNN model~\cite{arteaga2019deep} because it is designed to capture such graph structured mechanisms.
% We use the GCN model to learn the complex mechanism of cascading failures because the operational data of the power system is physically represented and correlated in the form of graphs.
% Compared with the DNN model~\cite{li2018alphago} and the CNN model~\cite{arteaga2019deep}, the proposed GCN model is designed to capture such graph structured mechanism.
The basic layers of the DNN model, the CNN model, and the GCN model are shown in \figurename~\ref{fig_layers}.
The above layers consist of a group of nodes.
The nodes in the next layer take the weighted summation of the nodes in the previous layer.
The layers mainly differ in the way that they take the weighted summation among different range of nodes in the previous layer.
The layers in DNN, CNN, and GCN take the weighted summation of all the nodes, the neighboring nodes in Euclidean domain, and the neighboring nodes in graph domain, respectively.
Benefiting from the structure of the graph convolutional layer, the GCN model can better capture the dependence and relationships represented by graphs.
% Hence, data that has the relationship represented by graphs can be better captured by the GCN model.

\subsection{Design of GCN Structure}

\begin{figure}[htb!]
	\centering
		\includegraphics[width=0.55\linewidth]{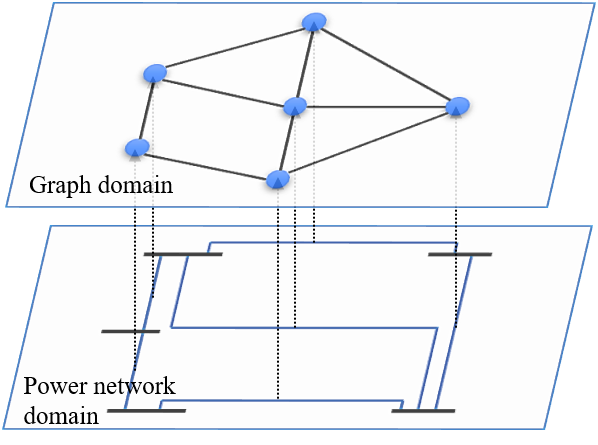}
	\caption{An illustration of the network-graph mapping, where branches in the power network are transformed to nodes in the graph.}
	\label{fig_branch2node}
\end{figure}
For clarity, we use ``bus'' and ``branch'' to represent the power network, while we use ``node'' and ``edge'' to represent the graph in GCN.
The traditional GCN models make classification or regression on nodes, but we search for the vulnerable branches in our framework.
Hence, we conduct a network-graph mapping as demonstrated in \figurename~\ref{fig_branch2node}, where the data organized in branches is transformed to the data organized in nodes.
% Hence, we transform the data organized in branches to the data organized in nodes, before feeding into the GCN network, as demonstrated in \figurename~\ref{fig_branch2node}.
The two nodes in the graph domain are connected when the two branches in the power network domain have common buses.
For a power network with $L$ branches, the graph domain has $L$ nodes.
The input of the GCN model can be formulated as:
\begin{equation}\label{eq_inputs}
  \bm{X}^{GCN}=\left[\bm{x}_1\text{ }...\text{ }\bm{x}_c\text{ }...\text{ }\bm{x}_C\right],
\end{equation}
where $C$ is the number of input features, $x_c$ is a $L\times 1$ vector.

\begin{figure*}[htb!]
	\centering
		\includegraphics[width=0.8\linewidth]{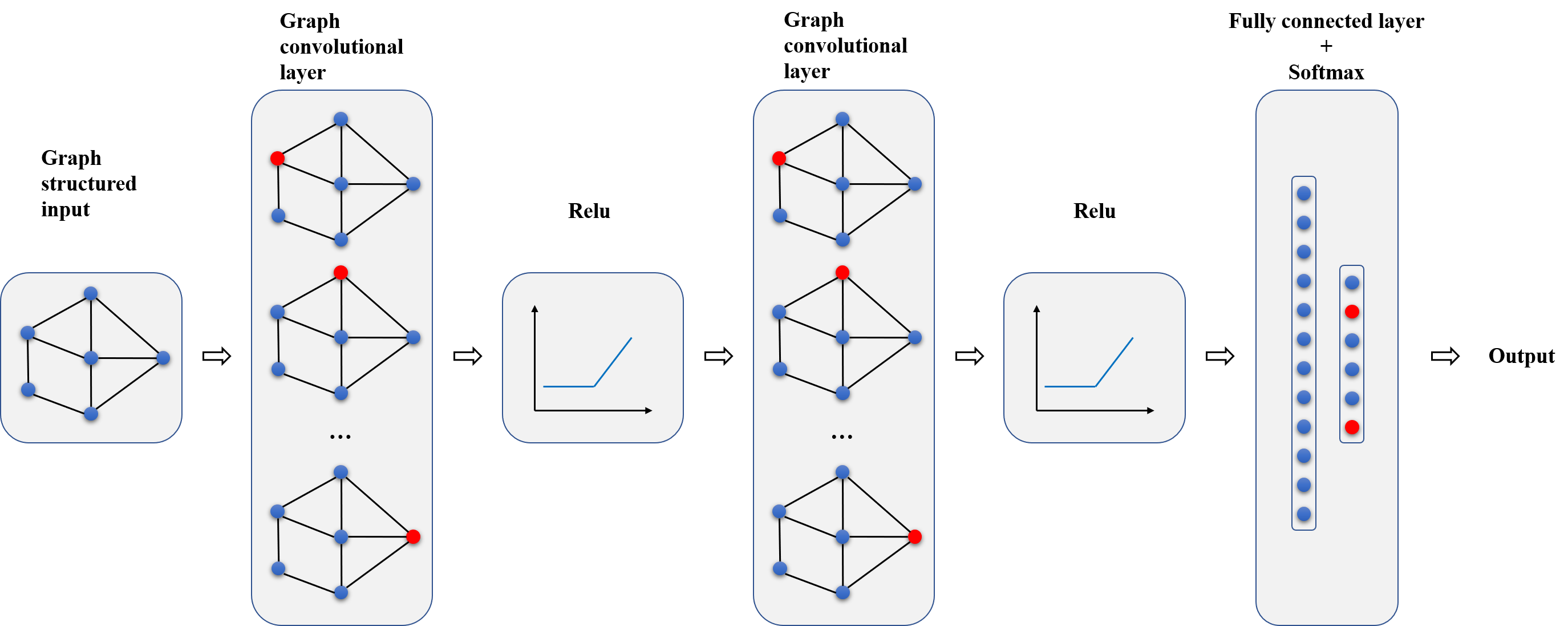}
	\caption{The structure of the proposed graph convolutional network model.}
	\label{fig_GCN_structure_whole}
\end{figure*}

The the GCN model to guide the cascading failure search is depicted in \figurename~\ref{fig_GCN_structure_whole}, which maps from $\bm{X}^{GCN}$ to $\bm{y}^{GCN}$.
It consists of two graph convolutional layers and a fully connected layer with softmax output.
The graph convolutional layer~\cite{du2017topology} takes the weighted average of the neighboring nodes in the graph domain and output a certain number of $L\times 1$ vectors (e.g. $F$ vectors).
The $f$th output vector is given by:
\begin{subequations}\label{eq_graph_convolutional}
  \begin{equation}\label{eq_graph_convolutional_a}
    \bm{y}_{f}^{graph}=\sum_{c=1}^{C} \mathbf{G}_{cf} \bm{x}_{c}+b_{f} \mathbf{1},
  \end{equation}
  \begin{equation}\label{eq_graph_convolutional_b}
    \mathbf{G}_{cf}=\sum_{k=0}^{K} w_{cfk} \mathbf{\overline{A}}^{k},
  \end{equation}
\end{subequations}
where $w_{cfk}$ is the learnable weight parameter, $b_{f}$ is the learnable constant parameter, $\mathbf{1}$ is the all-ones $L\times 1$ vector, $\mathbf{G}_{cf}$ is the $L\times L$ graph filter, $\mathbf{\overline{A}}=\mathbf{D}^{-1/2} \mathbf{A} \mathbf{D}^{-1/2}$ is the normalization of the graph adjacency matrix $\mathbf{A}$, and $\mathbf{D}$ is the diagonal degree matrix of the graph with $D_{ii}=\sum_{j}\mathbf{A}_{ij}$.
Equation~\eqref{eq_graph_convolutional_a} indicates that the output of the graph convolutional layer is the summation of inputs over graph filters.
Equation~\eqref{eq_graph_convolutional_b} indicates that the graph filter $\mathbf{G}_{cf}$ can extract the relationship within $K$ hops (or $K$ neighborhood) in a graph.

The rectified linear unit (Relu) layer serves as the activation function that add nonlinearity to the neural network.
It will output the positive part of the inputs:
\begin{equation}\label{eq_Relu}
  \bm{y}^{Relu}=\max (0,\bm{x}).
\end{equation}

The output dimension through the graph convolutional layer and the Relu layer is $L\times F$.
In one single node, the output can be transformed as a $F\times 1$ vector.
We add a fully connected layer:
\begin{equation}\label{eq_fully_connected}
  \bm{y}^{fully}=\mathbf{W}\bm{x}+b,
\end{equation}
where $\mathbf{W}$ is the learnable weights parameters, and $b$ is the learnable constant parameter.
In~\eqref{eq_fully_connected}, $\bm{y}^{fully}$ is a $L\times 2$ vector that consists of the variables representing the conditions for each node $k$: $y_{k1}^{fully}$ for normal condition and $y_{k2}^{fully}$ for load shedding.
The classification result is either normal condition or load shedding that with larger value of $y_{ki}^{fully}$.

We then introduce the softmax function to provide the probability of the final classification result:
\begin{equation}\label{eq_softmax}
  y_{ki}^{soft}=\frac{e^{y_{ki}^{fully}}}{e^{y_{k1}^{fully}}+e^{y_{k2}^{fully}}}\text{, }i=1,2.
\end{equation}
The classification error is set as the negative log-likelihood (NLL):
\begin{equation}\label{eq_nll_loss}
  L^{nll}_{k}\!=-\!w_1 y^{GCN}_k \!\log y_{k1}^{soft}\!-\!w_2 (1\!-\!y^{GCN}_k)\log y_{k2}^{soft},
\end{equation}
where $L^{nll}_{k}$ is the NLL of the $k$th output, $y^{GCN}_k$ is the ground truth output as introduced in~\eqref{eq_GCN}, and $w_1$ and $w_2$ are the weights to different classes.
As the output has less value-1 (load shedding) than value-0 (without load shedding), we assign $w_1>w_2$ to learn from biased sample.

\subsection{The Parameter Space of GCN}
We analyze the parameter space required to learn the cascading failure mechanism, by comparing the basic layers of the DNN model, the CNN model, and the GCN model again.
The comparison of the parameter space of the fully connected layer in the DNN model, the convolutional layer in the CNN model, and the graph convolutional layer in the GCN model are shown in Table~\ref{table_layers}. 
We assume the number of neurons of the fully connected layer is $N_{DNN}$, the kernel size of the convolutional layer is $K_{CNN}\times K_{CNN}$, and the number of hops of the graph convolutional layer is $K_{GCN}$.
See \cite{li2018alphago, arteaga2019deep} for modeling details of the DNN and the CNN, respectively.

\begin{table}[ht]
	\centering
	\caption{The Parameter Space of Three Types of Layers.}
	\label{table_layers}
		\begin{tabular}{@{}llll@{}}
      \toprule
         & Fully connected & Convolutional & Graph convolutional \\ \midrule
      Parameter space & $O(N_{DNN}^2)$           & $O(K_{CNN}^2)$          & $O(K_{GCN})$ \\
      \bottomrule
      \end{tabular}
	% }
\end{table}

In Table~\ref{table_layers}, $N_{DNN}$ is proportional to the number of buses in a power system. 
$K_{CNN}$ and $K_{GCN}$ denotes how many neighboring nodes to extract the feature from, which is far less than the number of $N_{DNN}$.
The neighboring nodes in convolutional layers may not be the neighboring buses in the power network, thus $N_{CNN}$ should be larger than $N_{GCN}$ to extract the feature of neighboring buses.
To sum up, regarding the cascading failure learning problem, the graph convolutional layer has the smallest feature space that is efficient to train.

\subsection{Interpretability of the Model}

% 有必要再说一下为什么要interpretable??

To increase the interpretability of the proposed GCN model, we firstly tailor the input of our model into four parts with physical interpretations, and then we calculate the contribution factors of the inputs by an LRP algorithm.

In detail, the input of the model is a $L\times 4$ matrix:
\begin{equation}\label{eq_total_inputs}
  \bm{X}^{GCN}=\left[\bm{x}^{t}\text{ }\bm{x}^{p}\text{ }\bm{x}^{b}\text{ }\bm{x}^{l}\right].
\end{equation}
In~\eqref{eq_total_inputs}, $\bm{x}^{t}$ is a boolean vector that represents the effects of current system topology (or the initial failures) on the cascading failures, with 0 for operational state and 1 for outage of a branch;
$\bm{x}^{p}$ is a vector that represents the effects of protection relays, the value is the ratio of absolute branch flow to the protection relay threshold branch flow (the $k$th value is $L_k/\beta L_k^{max}$);
$\bm{x}^{b}$ is the absolute branch flow vector;
and $\bm{x}^{l}$ represents the larger load value between the two buses that the branch connected to.
So far, we have organized our inputs into four aspects: topology, protection relay, branch flow, and the bus load.
% One could also add other factors that may be of possible interest.

To quantify the contribution factors of these four inputs, we implement the LRP algorithm that was proposed to explain the CNN model~\cite{bach2015pixel}.
The LRP decomposes the output value into a combination of the model inputs. 
The decomposed values, a.k.a. the relevance scores, quantify the contribution of the inputs to the output.
The process transforms the relevance scores of the next layer $R_i^{d+1}$ to the relevance scores of the previous layer $R_i^d$, such that the sum of each relevance scores of each layer equals to the output value:
\begin{equation}\label{eq_layers}
  y=\cdots=\!\!\!\!\sum_{i\in layer_{d+1}}\!\!\!\!R_i^{d+1}=\!\!\!\!\sum_{i\in layer_d}\!\!\!\!R_i^d=\cdots=\!\!\!\!\sum_{i\in layer_1}\!\!\!\!R_i^1.
\end{equation}
We decompose the relevance scores of the next layer by the weights of the positive inputs $z_{ij}^{+}$ from all the connected neurons of the previous layer:
\begin{equation}\label{eq_relevance}
  R_i^{d}=\sum_{j\in i}\frac{z_{ij}^{+}}{\sum_{i'\in j}z_{i'j}^{+}}R_j^{d+1},
\end{equation}
where $\sum_{j\in i}$ sums over the neurons in layer $d+1$ that connected with neuron $i$, $\sum_{i'\in j}$ sums over the neurons in layer $d$ that connected with neuron $j$, and $z_{ij}^{+}=\max(x_{i}w_{ij},0)$ is the positive part of the input from neuron $i$ to neuron $j$.
We decompose the relevance scores by the weight of positive parts to compute the excitatory effects on the output of each layer.
Recall that the Relu activation function only output the positive parts of the input, so that only positive input can contribute to larger output (or have excitatory effects) of the Relu activation function.
We only compute the excitatory effects to focus on the factors that incurs the load shedding.
In this work, we calculate the relevance scores corresponding to the output that determines the prediction of load shedding. 
For example, we calculate the factors that incurs the load shedding after the failure of branch~$k$.
In this case, according to the proposed GCN model, we decompose the output corresponding to the load shedding prediction, which is $y_{k1}^{fully}$.
That is, the $y$ in~\eqref{eq_layers} is $y_{k1}^{fully}$ in~\eqref{eq_fully_connected}.

% From~\eqref{eq_softmax}, $y_k^{GCN}$ is determined by $y_{k1}^{soft}-y_{k2}^{soft}$.
% The sensitivity of the $y_k^{GCN}$ to the inputs is given by the gradients $\nabla_{\bm{X}^{GCN}}(y_{k1}^{soft}-y_{k2}^{soft})$.
% The value of the gradients is computed by back propagation method~\cite{lecun2015deep}.
% Then, we clip the negative gradients and obtain the contribution factor matrix to the $k$th output:
% \begin{equation}\label{eq_contribution_factor}
%   \mathbf{C}_k=\max\left(\nabla_{\bm{X}^{GCN}}(y_{k1}^{soft}-y_{k2}^{soft}),0\right).
% \end{equation}
% The reason for clipping the negative gradients is to have a more sparse output so that we can concentrate on the factors that have excitatory effects on the load shedding.

% 可以从模型、其他研究、以及电网中的例子来说明为何取正部分

% The interpretability is a very recent research development for GCN, which is firstly proposed in
%这里应当是第二波亮点，要包含变量的物理意义，并且说明GCN的可解释性是very recent work
\section{Case Studies}
\label{section_case_studies}
\subsection{IEEE RTS-79 Test System}
The IEEE RTS-79 test system contains 24 buses and 38 branches with a peak load of 2850 MW and a total generation capacity of 3405 MW~\cite{subcommittee1979ieee}.
% The topology of the system is shown in \figurename~\ref{fig_RTS}.
8000 scenarios (the $\bm{X}^{GCN} \sim \bm{y}^{GCN}$ pair described in~\eqref{eq_GCN}) are generated as the training dataset, to consider the cases of different load profiles and different initial contingencies.
To model the load uncertainty, we multiply a random factor drawn uniformly over the interval [0.9, 1.1] to the load of every bus.
To simulate a more severe result, the actual load of each scenarios from~\cite{subcommittee1979ieee} is enlarged by a factor of 1.1.
In this work, we consider N-2 contingencies as the random initial failures.
The simulation of cascading failures is carried out based on MATLAB 2018b, Cplex~12.9~\cite{cplex}, and Matpower~6.0~\cite{zimmerman2010matpower}.
The proposed GCN model is implemented by Pytorch~1.3~\cite{paszke2019pytorch}.
The hyper-parameters of the GCN model are tuned by 20\% of the training data.
We set the number of hops $K=3$, the ratio of training weight $w_1/w_2=20$, and the number of the output vectors of the first and the second graph convolutional layer as $F_1=16$ and $F_2=4$, respectively.
The model is trained with 20 epochs and a batch size of 32, using the Adam optimizer~\cite{kingma2014adam} with the learning rate of 0.005.

% \begin{figure}[htb!]
% 	\centering
% 		\includegraphics[width=0.7\linewidth]{fig_RTS.png}
% 	\caption{The topology of the IEEE RTS-79 system.}
% 	\label{fig_RTS}
% \end{figure}

\begin{figure}[htb!]
	\centering
		\includegraphics[width=0.8\linewidth]{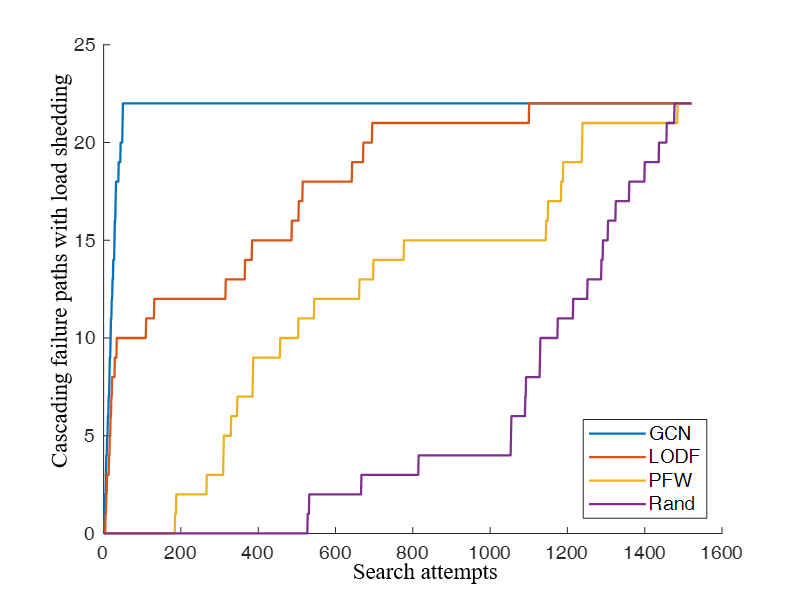}
	\caption{The number of detected cascading failure paths that incurs load shedding as a function of search attempts: the case of IEEE RTS-79 test system.}
	\label{fig_search_RTS}
\end{figure}

We implement a set of newly generated load profiles to test different cascading failure online search methods. 
Each compared method only differs from the \textbf{Algorithm~\ref{algorithm_online_search}} in the orders to disconnect the branches in \textbf{Step~2*}.
\begin{enumerate}[1)]
\item \textbf{Rand}: The branches are disconnected randomly;
\item \textbf{PFW}: The branches are disconnected from the largest power flow value to the smallest power flow value. It is the deterministic form of the power flow weighted method proposed in~\cite{zhang2019online}; %It only differs from the \textbf{Algorithm~\ref{algorithm_online_search}} in the orders to disconnect the branches in \textbf{Step~2*}. 
\item \textbf{LODF}: The branches are disconnected by the value of $\bm{y}^{P}$ described in~\eqref{eq_LODF}-\eqref{eq_vul_physical}, from the largest value to the smallest value. The LODF-based method and its derivatives are widely used to search the cascading failures~\cite{yao2016risk,soltan2017analyzing}
\item \textbf{GCN}: \textbf{Algorithm~\ref{algorithm_online_search}} proposed in this work.
\end{enumerate}
We present the number of detected cascading failure paths that incurs load shedding in \figurename~\ref{fig_search_RTS}.
After exhaustive search, there are 22 cascading failure paths that incur load shedding.
The proposed \textbf{GCN} method can detect all the paths after only 50 search attempts, while method \textbf{LODF}, \textbf{PFW}, and \textbf{Rand} detect all the paths after 1101, 1485, and 1477 search attempts, respectively.
Hence, the proposed \textbf{GCN} method has clearly higher search efficiency than all other compared methods.

\begin{figure}[htb!]
  \begin{minipage}{9cm}
    \centering
    \includegraphics[width=0.95\linewidth]{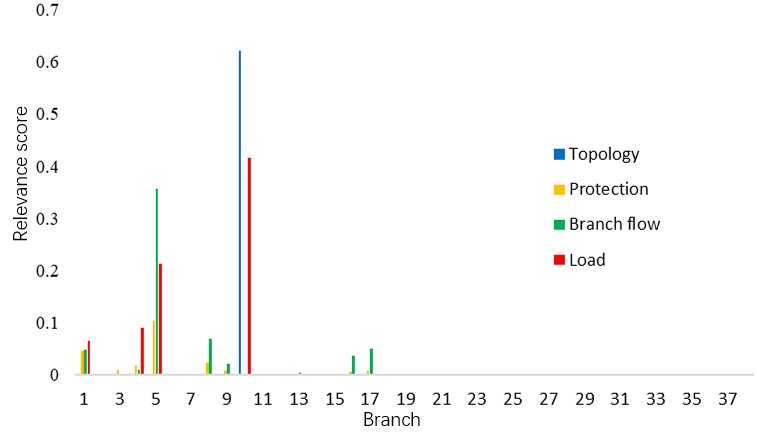}
    \centerline{\footnotesize{(a)}}
  \end{minipage}
  \begin{minipage}{9cm}
    \centering
    \includegraphics[width=1\linewidth]{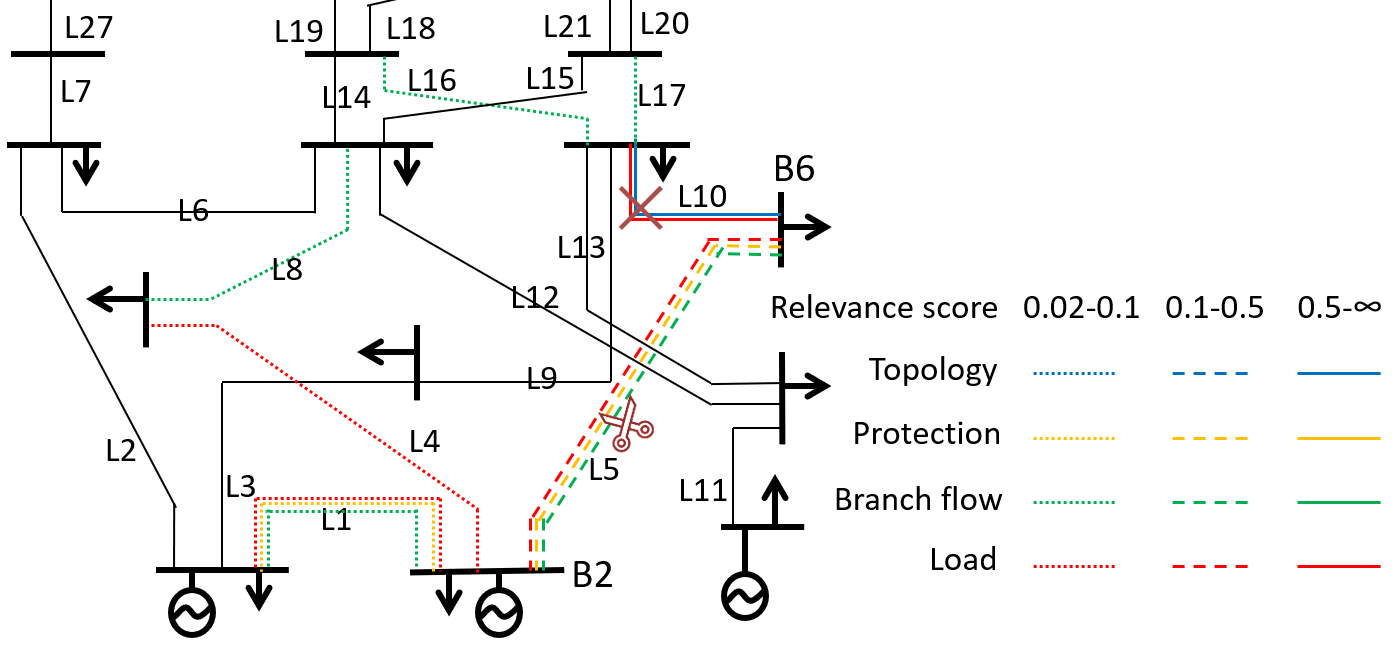}
    \centerline{\footnotesize{(b)}}
  \end{minipage}
	\caption{The relevance scores for each inputs of the GCN model that triggers the load shedding for the initial failures L10$\rightarrow$L5. (a) The bar graph. (b) The topological graph that represents a part of the IEEE RTS-79 system.}
	\label{fig_inter_L10-5}
\end{figure}

We then show the interpretability of the model by the proposed LRP method, which present the relevance scores of the inputs to a certain output.
The relevance scores are the contribution factors of the inputs to the load shedding events.
We use `L' for branch and `B' for bus.
From the above 22 cascading failure paths, we analyze the case with initial failures L10$\rightarrow$L5.
That is, we show the relevance scores to the 5th output of vector $\bm{y}^{GCN}$, under the power system states where L10 is already disconnected, as is depicted in \figurename~\ref{fig_inter_L10-5}.
A direct physical explanation of the load shedding is that the disconnection of L10 and L5 will result in the islanding and load shedding of B6.
The relevance scores in \figurename~\ref{fig_inter_L10-5} can well explain the above inference.
The load shedding in this case mainly results from the initial failure of L10 (the topology of L10), the increasing load of B6 (the load corresponding to L5 and L10), and the increasing branch flow from B2 to B6 (the branch flow of L5).
Note that some factors are not directly related to the load shedding, but they are related to the factors that directly cause the load shedding.
For example, the protection of L5 is not directly related to the load shedding of B6, but the increase of the branch flow / protection threshold ratio is closely related with the branch flow of L5.
We cannot simply remove the protection vector $\bm{x}^p$ from the inputs, because it will serve as an independent determinant as will be shown in the next case.

\begin{figure}[htb!]
  % \begin{minipage}{9cm}
  %   \centering
  %   \includegraphics[width=0.95\linewidth]{fig_interpretability_L27-2.png}
  %   \centerline{\footnotesize{(a)}}
  % \end{minipage}
  % \begin{minipage}{9cm}
  %   \centering
  %   \includegraphics[width=0.79\linewidth]{fig_interpretability_L27-2_topo.png}
  %   \centerline{\footnotesize{(b)}}
  % \end{minipage}
  \centering
  \includegraphics[width=0.81\linewidth]{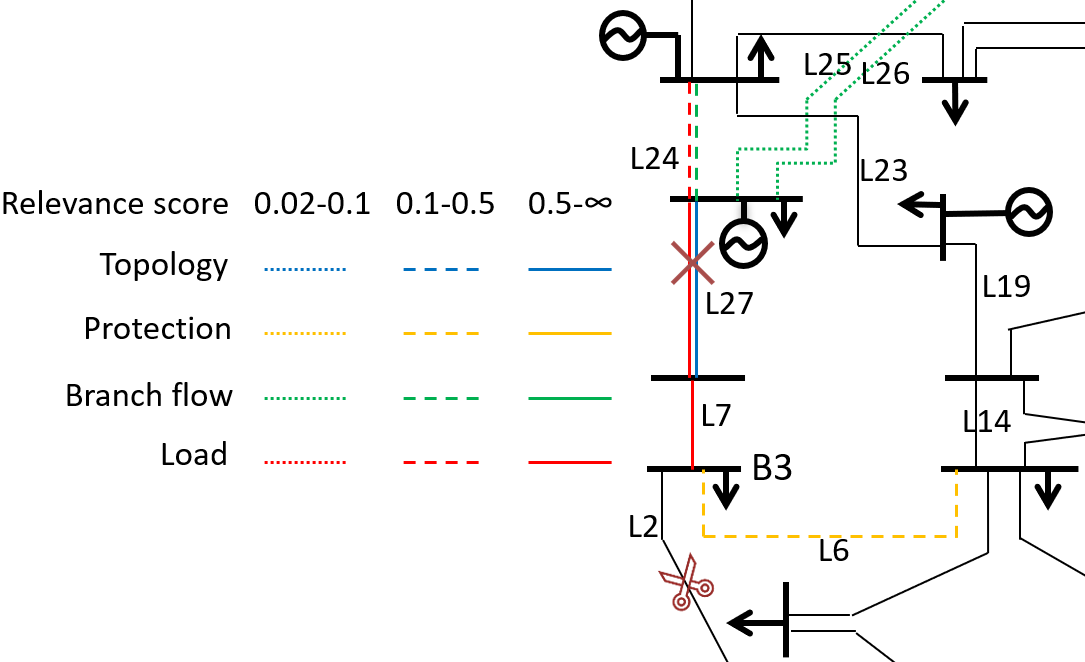}
	\caption{The relevance scores for each inputs of the GCN model that triggers the load shedding for the initial failures L27$\rightarrow$L2. The topology is only a part of the IEEE RTS-79 system.}% (a) The bar graph. (b) The topological graph.
	\label{fig_inter_L27-2}
\end{figure}

The above interpretation results can be directly inferred from the topology of the system.
To further demonstrate the interpretability of the LRP algorithm, we analyze a cascading failures case that is less apparent.
The relevance scores to the 2nd output of vector $\bm{y}^{GCN}$, under the power system states that L27 is already disconnected (L27$\rightarrow$L2), is depicted in \figurename~\ref{fig_inter_L27-2}.
The disconnection of L27 and L2 will not trigger any islanding conditions, but the load in B3 relies only on the transmission line of L6.
The branch flow of L6 will exceed the maximum value. 
Subsequently, the re-dispatching process will decrease the branch flow of L6 and causes the load shedding of B3.
The relevance scores in \figurename~\ref{fig_inter_L27-2} can also well explain this process.
The load shedding in this case mainly results from the initial failure of L27 (the topology of L27), the increasing load of B3 (the load corresponding to L7), and the branch flow limits of L6 (the protection of L6).
The above cases show that the major contribution factors of the prediction results correspond to the physical inference of the cascading failures.
The interpretation method can explore some less obvious factors (e.g. the protection of L6 in \figurename~\ref{fig_inter_L27-2}) that incurs the load shedding.
Still, the contribution factors also includes some indirect factors (e.g. the protection of L5 in \figurename~\ref{fig_inter_L10-5}), which suggests that operators should be more careful or should update the model when the operation pattern changes.

\subsection{The Power System of Henan Province}
The power system of Henan province considered in this work is a 500kV-220kV power system.
After external network equivalence, the power system contains 730 buses and 1571 branches.
To consider a more severe circumstance, we enlarge the actual load profiles of Henan province by a factor of 1.1, and we reduce all the maximum branch flow thresholds by a factor of 0.8.
Since the load shedding events are more rare in the Henan province case, we increase the ratio of training weight to $w_1/w_2=800$.
50000 scenarios are generated as the training dataset to model different load profiles and different initial contingencies.
The model is trained with 40 epochs and a batch size of 256.
All other experiment settings, including the cascading failures data generation and the model training processes, are the same with the IEEE RTS-79 test system.
% 如何仿真采样真的是一个很重要的问题！！毕竟人家很多paper都说到了

We first use the power system of Henan province to test the prediction accuracy of the proposed GCN model, compared to the DNN and CNN model mentioned above.
We generate 10000 new scenarios as the testing dataset. 
The DNN, CNN, and the GCN models are trained and tested under the same training and testing datasets, respectively.
The DNN model has three fully connected layers with $4L$, $L$, and $L$ output neurons, respectively.
The CNN model has two convolutional layers with one fully connected layer.
The kernel size of the two convolutional layers are $3\times 3\times 16$ and $3\times 3\times 4$, respectively.
All of the models implement the Relu activation function.
Since the number of samples from different classes is unbalanced, a high total prediction accuracy may not be conducive to an efficient search of cascading failures.
We therefore implement different indexes to evaluate the prediction result.
We first introduce the confusion matrix in Table~\ref{table_confusion}.

\begin{table}[htb]
  \centering
	\caption{The confusion matrix.}
  \label{table_confusion}
  \setlength{\tabcolsep}{1.1mm}{
    \begin{tabular}{@{}llll@{}}
      \toprule
      \multicolumn{2}{l}{\multirow{2}{*}{}}&\multicolumn{2}{l}{Predicted results} \\ 
      \cmidrule(l){3-4} \multicolumn{2}{l}{}&without load shedding&load shedding \\ 
      \midrule
      \multirow{2}{*}{Actual results} & without load shedding & a & b\\ 
      \cmidrule(l){2-4} & load shedding & c & d \\ \bottomrule
    \end{tabular}
  }
\end{table}

Several indexes are introduced~\cite{powers2011evaluation}:
\begin{enumerate}[1)]
  \item Total accuracy: $(a+d)/(a+b+c+d)$;
  \item Hit rate: $d/(b+d)$; %the ratio of actual predictions when the prediction result is load shedding, 
  \item Cover rate: $d/(c+d)$; % the ratio of predicted load shedding events from all the load shedding events, 
\end{enumerate}

\begin{table}[htb]
  \centering
	\caption{The comparison of DNN, CNN, and GCN models.}
  \label{table_comparison_models}
  \setlength{\tabcolsep}{4.1mm}{
  \begin{tabular}{@{}llll@{}}
    \toprule
               & DNN    & CNN    & GCN    \\ \midrule
    Total accuracy   & 0.9416 & 0.9844 & 0.9988 \\
    Hit rate   & 0.0080 & 0.0279 & 0.3013 \\
    Cover rate & 0.9460 & 0.8963 & 0.9961 \\ \bottomrule
  \end{tabular}
  }
\end{table}

The calculated values of the above indexes of DNN, CNN, and GCN model are listed in Table~\ref{table_comparison_models}.
Note that the prediction of the load shedding events is not the final result, but acts as a guidance for the search for cascading failures that will be implemented by physical models, as shown in \textbf{Algorithm~\ref{algorithm_online_search}}.
The total accuracy reflects the prediction results of both the cases with and without load shedding.
In practice, however, the prediction of load shedding events is more related to the search efficiency of cascading failures.
The hit rate measures how well the predicted load shedding events hit the actual load shedding.
The cover rate is the ratio of predicted load shedding events to all the load shedding events.
A small hit rate will result in useless search attempts, while a small cover rate will miss some cascading failure events so that many additional search attempts are required to compensate for the missing events.
From Table~\ref{table_comparison_models}, the proposed GCN model has the best performance not only in the total accuracy, but also in terms of the hit rate and the cover rate.

\begin{figure}[htb!]
	\centering
		\includegraphics[width=0.8\linewidth]{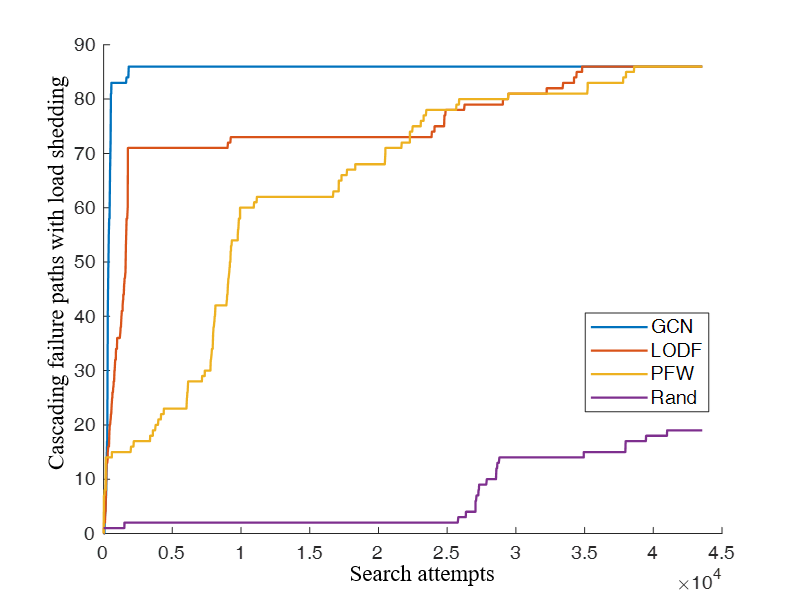}
	\caption{The number of detected cascading failure paths that incurs load shedding as a function of search attempts: the case of Henan provence power system.}
	\label{fig_search_Henan}
\end{figure}

Next, we present the search efficiency on a set of new load profiles of the power system of Henan province.
The search performance of method \textbf{Rand}, \textbf{PFW}, \textbf{LODF}, and \textbf{GCN} are shown in \figurename~\ref{fig_search_Henan}.
An exhaustive search identifies 86 cascading failure paths that incur load shedding.
The proposed \textbf{GCN} method can detect all the paths after 2160 search attempts, while methods \textbf{LODF} and \textbf{PFW} detect all the paths after $34822$ and $38608$ search attempts, respectively.
Method \textbf{Rand} can only find a small proportion of all the paths in the given search attempts.
The \textbf{LODF} and the \textbf{PFW} method have a high search efficiency at the first few search attempts, but require a high number of search attempts to detect the remaining cascading failure paths that incur load shedding.
This is because some of the load shedding events can be easily detected by the physical rules (e.g. the islanding conditions), while the rest of the events are complex and cannot be detected by explicit physical rules.
The above results indicate that the proposed GCN based method has a high efficiency in searching for the load shedding events caused by cascading failures.

\section{Conclusions}
\label{section_conclusions}
Our work provides a GCN based approach to conduct online search for the power system cascading failures with high efficiency.
The proposed method links the power network topology with the structure of neural networks and therefore yeilds a smaller parameter space to predict the load shedding events caused by cascading failures.
We also provide an LRP algorithm to uncover the reasons for predicting the cascading failure events.
Case studies on the IEEE RTS-79 test system and on China's Henan province power system validate the superiority of the method: 
1) the GCN model has a better prediction accuracy than benchmark neural network models;
2) the GCN based cascading search method has a higher efficiency in detecting the load shedding events caused by cascading failures;
3) the prediction results of the GCN model are interpretable, enabling the user to check the logic of the model and uncover the factors that cause the cascading failures.
Overall, the proposed method enables an efficient and interpretable online cascading failure search that can contribute to a reliable operation of power systems.

%-----------------------------------------
%-----------------------------------------
%-----------------------------------------

\ifCLASSOPTIONcaptionsoff
  \newpage
\fi


% \bibliographystyle{IEEEtran}
%\balance
\bibliography{IEEEabrv,myReference}

\begin{thebibliography}{10}
\providecommand{\url}[1]{#1}
\csname url@samestyle\endcsname
\providecommand{\newblock}{\relax}
\providecommand{\bibinfo}[2]{#2}
\providecommand{\BIBentrySTDinterwordspacing}{\spaceskip=0pt\relax}
\providecommand{\BIBentryALTinterwordstretchfactor}{4}
\providecommand{\BIBentryALTinterwordspacing}{\spaceskip=\fontdimen2\font plus
\BIBentryALTinterwordstretchfactor\fontdimen3\font minus
  \fontdimen4\font\relax}
\providecommand{\BIBforeignlanguage}[2]{{%
\expandafter\ifx\csname l@#1\endcsname\relax
\typeout{** WARNING: IEEEtran.bst: No hyphenation pattern has been}%
\typeout{** loaded for the language `#1'. Using the pattern for}%
\typeout{** the default language instead.}%
\else
\language=\csname l@#1\endcsname
\fi
#2}}
\providecommand{\BIBdecl}{\relax}
\BIBdecl

\bibitem{vaiman2012risk}
M.~Vaiman, K.~Bell, Y.~Chen, B.~Chowdhury, I.~Dobson, P.~Hines, M.~Papic,
  S.~Miller, and P.~Zhang, ``Risk assessment of cascading outages:
  Methodologies and challenges,'' \emph{IEEE Transactions on Power Systems},
  vol.~27, no.~2, p. 631, 2012.

\bibitem{bienstock2015electrical}
D.~Bienstock, \emph{Electrical transmission system cascades and vulnerability:
  an operations research viewpoint}.\hskip 1em plus 0.5em minus 0.4em\relax
  SIAM, 2015, vol.~22.

\bibitem{haes2019survey}
H.~Haes~Alhelou, M.~E. Hamedani-Golshan, T.~C. Njenda, and P.~Siano, ``A survey
  on power system blackout and cascading events: Research motivations and
  challenges,'' \emph{Energies}, vol.~12, no.~4, p. 682, 2019.

\bibitem{athari2017impacts}
M.~H. Athari and Z.~Wang, ``Impacts of wind power uncertainty on grid
  vulnerability to cascading overload failures,'' \emph{IEEE Transactions on
  Sustainable Energy}, vol.~9, no.~1, pp. 128--137, 2017.

\bibitem{chen2013composite}
Q.~Chen and L.~Mili, ``Composite power system vulnerability evaluation to
  cascading failures using importance sampling and antithetic variates,''
  \emph{IEEE transactions on power systems}, vol.~28, no.~3, pp. 2321--2330,
  2013.

\bibitem{henneaux2014improving}
P.~Henneaux and P.-E. Labeau, ``Improving monte carlo simulation efficiency of
  level-i blackout probabilistic risk assessment,'' in \emph{2014 International
  Conference on Probabilistic Methods Applied to Power Systems (PMAPS)}.\hskip
  1em plus 0.5em minus 0.4em\relax IEEE, 2014, pp. 1--6.

\bibitem{kim2012splitting}
J.~Kim, J.~A. Bucklew, and I.~Dobson, ``Splitting method for speedy simulation
  of cascading blackouts,'' \emph{IEEE Transactions on Power Systems}, vol.~28,
  no.~3, pp. 3010--3017, 2012.

\bibitem{wang2014efficient}
S.-P. Wang, A.~Chen, C.-W. Liu, C.-H. Chen, J.~Shortle, and J.-Y. Wu,
  ``Efficient splitting simulation for blackout analysis,'' \emph{IEEE
  Transactions on Power Systems}, vol.~30, no.~4, pp. 1775--1783, 2014.

\bibitem{eppstein2012random}
M.~J. Eppstein and P.~D. Hines, ``A “random chemistry” algorithm for
  identifying collections of multiple contingencies that initiate cascading
  failure,'' \emph{IEEE Transactions on Power Systems}, vol.~27, no.~3, pp.
  1698--1705, 2012.

\bibitem{yao2016risk}
R.~Yao, S.~Huang, K.~Sun, F.~Liu, X.~Zhang, S.~Mei, W.~Wei, and L.~Ding, ``Risk
  assessment of multi-timescale cascading outages based on markovian tree
  search,'' \emph{IEEE Transactions on Power Systems}, vol.~32, no.~4, pp.
  2887--2900, 2016.

\bibitem{liu2019fast}
Y.~Liu, Y.~Wang, P.~Yong, N.~Zhang, C.~Kang, and D.~Lu, ``Fast power system
  cascading failure path searching with high wind power penetration,''
  \emph{IEEE Transactions on Sustainable Energy}, 2019.

\bibitem{soltan2015analysis}
S.~Soltan, D.~Mazauric, and G.~Zussman, ``Analysis of failures in power
  grids,'' \emph{IEEE Transactions on Control of Network Systems}, vol.~4,
  no.~2, pp. 288--300, 2015.

\bibitem{soltan2017analyzing}
S.~Soltan, A.~Loh, and G.~Zussman, ``Analyzing and quantifying the effect of $
  k $-line failures in power grids,'' \emph{IEEE Transactions on Control of
  Network Systems}, vol.~5, no.~3, pp. 1424--1433, 2017.

\bibitem{li2018alphago}
F.~Li and Y.~Du, ``From alphago to power system ai: What engineers can learn
  from solving the most complex board game,'' \emph{IEEE Power and Energy
  Magazine}, vol.~16, no.~2, pp. 76--84, 2018.

\bibitem{du2019achieving}
Y.~Du, F.~F. Li, J.~Li, and T.~Zheng, ``Achieving 100x acceleration for n-1
  contingency screening with uncertain scenarios using deep convolutional
  neural network,'' \emph{IEEE Transactions on Power Systems}, 2019.

\bibitem{arteaga2019deep}
J.-M.~H. Arteaga, F.~Hancharou, F.~Thams, and S.~Chatzivasileiadis, ``Deep
  learning for power system security assessment,'' in \emph{13th IEEE PowerTech
  2019}.\hskip 1em plus 0.5em minus 0.4em\relax IEEE, 2019.

\bibitem{sun2018deep}
M.~Sun, I.~Konstantelos, and G.~Strbac, ``A deep learning-based feature
  extraction framework for system security assessment,'' \emph{IEEE
  Transactions on Smart Grid}, 2018.

\bibitem{yan2016q}
J.~Yan, H.~He, X.~Zhong, and Y.~Tang, ``Q-learning-based vulnerability analysis
  of smart grid against sequential topology attacks,'' \emph{IEEE Transactions
  on Information Forensics and Security}, vol.~12, no.~1, pp. 200--210, 2016.

\bibitem{zhang2019online}
Z.~Zhang, S.~Huang, Y.~Chen, S.~Mei, R.~Yao, and K.~Sun, ``An online search
  method for representative risky fault chains based on reinforcement learning
  and knowledge transfer,'' \emph{IEEE Transactions on Power Systems}, 2019.

\bibitem{hines2016cascading}
P.~D. Hines, I.~Dobson, and P.~Rezaei, ``Cascading power outages propagate
  locally in an influence graph that is not the actual grid topology,''
  \emph{IEEE Transactions on Power Systems}, vol.~32, no.~2, pp. 958--967,
  2016.

\bibitem{cremer2019optimization}
J.~Cremer, I.~Konstantelos, and G.~Strbac, ``From optimization-based machine
  learning to interpretable security rules for operation,'' \emph{IEEE
  Transactions on Power Systems}, 2019.

\bibitem{wu2019comprehensive}
Z.~Wu, S.~Pan, F.~Chen, G.~Long, C.~Zhang, and P.~S. Yu, ``A comprehensive
  survey on graph neural networks,'' \emph{arXiv preprint arXiv:1901.00596},
  2019.

\bibitem{du2017topology}
J.~Du, S.~Zhang, G.~Wu, J.~M. Moura, and S.~Kar, ``Topology adaptive graph
  convolutional networks,'' \emph{arXiv preprint arXiv:1710.10370}, 2017.

\bibitem{baldassarre2019explainability}
F.~Baldassarre and H.~Azizpour, ``Explainability techniques for graph
  convolutional networks,'' \emph{arXiv preprint arXiv:1905.13686}, 2019.

\bibitem{carreras2004complex}
B.~A. Carreras, V.~E. Lynch, I.~Dobson, and D.~E. Newman, ``Complex dynamics of
  blackouts in power transmission systems,'' \emph{Chaos: An Interdisciplinary
  Journal of Nonlinear Science}, vol.~14, no.~3, pp. 643--652, 2004.

\bibitem{ren2008long}
H.~Ren, I.~Dobson, and B.~A. Carreras, ``Long-term effect of the n-1 criterion
  on cascading line outages in an evolving power transmission grid,''
  \emph{IEEE transactions on power systems}, vol.~23, no.~3, pp. 1217--1225,
  2008.

\bibitem{mei2009improved}
S.~Mei, F.~He, X.~Zhang, S.~Wu, and G.~Wang, ``An improved opa model and
  blackout risk assessment,'' \emph{IEEE Transactions on Power Systems},
  vol.~24, no.~2, pp. 814--823, 2009.

\bibitem{guler2007generalized}
T.~Guler, G.~Gross, and M.~Liu, ``Generalized line outage distribution
  factors,'' \emph{IEEE Transactions on Power Systems}, vol.~22, no.~2, pp.
  879--881, 2007.

\bibitem{tejada2017security}
D.~A. Tejada-Arango, P.~S{\'a}nchez-Mart{\i}n, and A.~Ramos, ``Security
  constrained unit commitment using line outage distribution factors,''
  \emph{IEEE Transactions on Power Systems}, vol.~33, no.~1, pp. 329--337,
  2017.

\bibitem{bach2015pixel}
S.~Bach, A.~Binder, G.~Montavon, F.~Klauschen, K.-R. M{\"u}ller, and W.~Samek,
  ``On pixel-wise explanations for non-linear classifier decisions by
  layer-wise relevance propagation,'' \emph{PloS one}, vol.~10, no.~7, p.
  e0130140, 2015.

\bibitem{subcommittee1979ieee}
{Probability Methods Subcommittee}, ``{IEEE} reliability test system,''
  \emph{IEEE Trans. power apparatus and systems}, no.~6, pp. 2047--2054, 1979.

\bibitem{cplex}
T.~S. Manuals, ``Cplex document,''
  \url{https://www.ibm.com/products/ilog-cplex-optimization-studio}, 2019.

\bibitem{zimmerman2010matpower}
R.~D. Zimmerman, C.~E. Murillo-S{\'a}nchez, and R.~J. Thomas, ``Matpower:
  Steady-state operations, planning, and analysis tools for power systems
  research and education,'' \emph{IEEE Transactions on power systems}, vol.~26,
  no.~1, pp. 12--19, 2010.

\bibitem{paszke2019pytorch}
A.~Paszke, S.~Gross, F.~Massa, A.~Lerer, J.~Bradbury, G.~Chanan, T.~Killeen,
  Z.~Lin, N.~Gimelshein, L.~Antiga \emph{et~al.}, ``Pytorch: An imperative
  style, high-performance deep learning library,'' in \emph{Advances in Neural
  Information Processing Systems}, 2019, pp. 8024--8035.

\bibitem{kingma2014adam}
D.~P. Kingma and J.~Ba, ``Adam: A method for stochastic optimization,''
  \emph{arXiv preprint arXiv:1412.6980}, 2014.

\bibitem{powers2011evaluation}
D.~M. Powers, ``Evaluation: from precision, recall and f-measure to roc,
  informedness, markedness and correlation,'' 2011.

\end{thebibliography}
%\bibliography{myReference} 



%\end{IEEEbiography}
%% insert where needed to balance the two columns on the last page with
%%\newpage


%\vfill


\end{document}